\documentclass[]{jkas}

\def\year{2025} 
\def\volume{00} 
\def\issue{0} 
\def\beginpage{231} 
\def\endpage{242} 
\def\received{July 16, 2025} 
\def\accepted{September 27, 2025} 
\def\published{October 10, 2025}
\date{Received \received; Accepted \accepted; Published \published}

\newcommand{\Msun}{\rm M_{\odot}}

\title{YSO Variability in the W51 Star-Forming Region}

\author[1]{Mi-Ryang Kim}{0000-0002-1408-7747} 
\author[1,2]{Jeong-Eun Lee}{0000-0003-3119-2087}
\author[1,3]{Contreras Pe{\~n}a Carlos}{0000-0003-1894-1880}
\author[4,5]{Gregory Herczeg}{0000-0002-7154-6065}
\author[6,7]{Doug Johnstone}{0000-0002-6773-459X}
\author[8]{Miju Kang}{0000-0002-5016-050X}

\affil[1]{Astronomy Program, Department of Physics and Astronomy, Seoul National University, Seoul 08826, Republic of Korea}
\affil[2]{SNU Astronomy Research Center, Seoul National University, Seoul 08826, Republic of Korea}
\affil[3]{Research Institute of Basic Sciences, Seoul National University, Seoul 08826, Republic of Korea}

\affil[4]{Kavli Institute for Astronomy and Astrophysics, Peking University, Yiheyuan Lu 5, Haidian Qu, 100871 Beijing, People's Republic of China}
\affil[5]{Department of Astronomy, Peking University, Yiheyuan 5, Haidian Qu, 100871 Beijing, People's Republic of China}

\affil[6]{NRC Herzberg Astronomy and Astrophysics, 5071 West Saanich Rd, Victoria, BC, V9E 2E7, Canada}
\affil[7]{Department of Physics and Astronomy, University of Victoria, Victoria, BC, V8P 5C2, Canada}

\affil[8]{Korea Astronomy and Space Science Institute, Daejeon 34055, Republic of Korea}

\def\corrauthor{Mi-Ryang Kim, \email{mi.ryang@snu.ac.kr} and Jeong-Eun Lee, \email{lee.jeongeun@snu.ac.kr}}
\def\runningauthor{M. Kim et al.}
\def\runningtitle{W51}

\def\keywords{stars: formation — stars: pre-main-sequence — stars: protostars — stars: variables: T Tauri, Herbig Ae/Be — catalogues}

\def\abstracttext{
Time-domain statistical studies of mid-infrared and submm variability in nearby star-forming regions show that at least half of all protostars are variable. In this study, we present a statistical analysis of the mid-infrared variability of Young Stellar Objects (YSOs) in the distant, massive star-forming region W51, based on data from the Near-Earth Object Wide-field Infrared Survey Explorer (NEOWISE) All-Sky Survey. From a catalog of 81 protostars, 527 disk objects, and 37,687 other sources, including diskless pre-main sequence stars (Class III) and likely evolved contaminants such as AGB stars (collectively labeled as PMS+E), we identified significant variability in both the 3.4 $\mu$m (W1) and 4.6 $\mu$m (W2) bands. Due to the large distance ($\sim$5.4 kpc) and high extinction of W51, our sample primarily includes intermediate- to high-mass YSOs ($\geq$2 $\Msun$), in contrast to nearby star-forming regions such as Taurus or the Gould Belt, which are dominated by low-mass stars ($\leq$1 $\Msun$). This mass bias may affect the observed variability characteristics and their interpretation. Specifically, 11.1\% of protostars, 7.6\% of Disk objects, and 0.6\% of PMS+E sources exhibited secular variability in the W2, while 8.6\% of protostars, 2.3\% of Disk objects, and 0.5\% of PMS+E sources exhibited stochastic variability. Similar trends were observed in the W1 band.
Both the fraction and amplitude of variability increase toward earlier evolutionary stages. Protostars exhibit predominantly stochastic variability with high amplitudes, likely driven by dynamic accretion, and extinction changes. In contrast, disk objects show more secular variability, including linear, curved, and periodic patterns, possibly caused by moderate accretion changes or geometric modulation in the inner disk. Analysis of brightness and color changes revealed that protostars typically become redder as they brighten, while disk objects show more complex behavior: they appear roughly balanced in W1 but more often become bluer in W2. These trends are consistent with enhanced dust emission or extinction in protostars, and reduced extinction or accretion-induced hotspot modulation in disk objects. These trends reflect the distinct physical mechanisms at play across evolutionary stages and demonstrate that mid-infrared variability provides useful insight into the accretion and disk evolution processes in young stars.}

\begin{document}
\jkashead

\section{Introduction}
\label{variability}
Stars form as molecular clouds undergo gravitational collapse, giving rise to protostars that accrete material through circumstellar disks. While the formation and evolution of low-mass YSOs have been extensively studied and are relatively well understood, the physical processes governing high-mass star formation remain less constrained due to their short evolutionary timescales, large distances, and complex environments \citep{Zinnecker.2007, Tan.2014, Motte.2018}.
Since high-mass YSOs are often deeply embedded and distant, studying their variability can give useful clues about their inner structure and evolution.
Observations across multiple wavelengths have revealed diverse patterns of variability in young stellar objects (YSOs), from short-term flickering over hours to long-term brightness trends spanning several years. These variations trace a range of physical processes, including dynamic accretion flows, variable extinction, disk structure evolution, and surface phenomena such as hot or cool spots \citep[see review by][]{Fischer.2023}.
Long-term monitoring in the mid-infrared has proven especially useful for probing embedded protostars, revealing both stochastic and secular variability associated with changes in accretion luminosity and circumstellar dust geometry \citep{Scholz.2013, Carlos.2020, Zakri.2022,Carlos.2025}.

Previous studies of low-mass star-forming regions have highlighted the importance of variability in YSOs across multiple wavelengths. For example, \citet{ParkWS.2021} identified about 1700 variable YSOs using NEOWISE mid-infrared data and classified them into secular and stochastic types based on their 4.6 $\mu$m light curves, finding that earlier-stage YSOs exhibit more secular variability with higher amplitudes. 
A recent large-scale analysis by \citet{Neha.2025} extended this framework to over 20,000 YSO candidates from the SPICY catalog \citep{Kuhn.2021}, identifying variability in more than 5,000 sources and confirming that Class I objects show the highest variability fraction, with redder-when-brighter and bluer-when-brighter trends linked to different variability mechanisms. In addition, \citet{ParkGS.2024} and \citet{Chen.2025} studied YSOs in the M17 region with NEOWISE and the JCMT Transient Survey, showing that mid-IR variability is strongly concentrated among Class 0/I objects, while one submillimeter-brightening protostar exhibited a dramatic luminosity increase of over 16\%. These findings are consistent with other mid-IR studies of embedded protostars \citep{Scholz.2013, Wolk.2018, Zakri.2022}. 
Complementary near-infrared surveys further support this picture. Large-area variability searches, such as the UKIDSS Galactic Plane Survey \citep[e.g.,][]{Lucas.2017}, have uncovered hundreds of high-amplitude infrared variables, the majority of which are YSOs. These studies confirmed that accretion-driven processes dominate large-amplitude NIR variability, consistent with mid-IR results.

\begin{figure}
    \centering
    \includegraphics[width=\columnwidth]{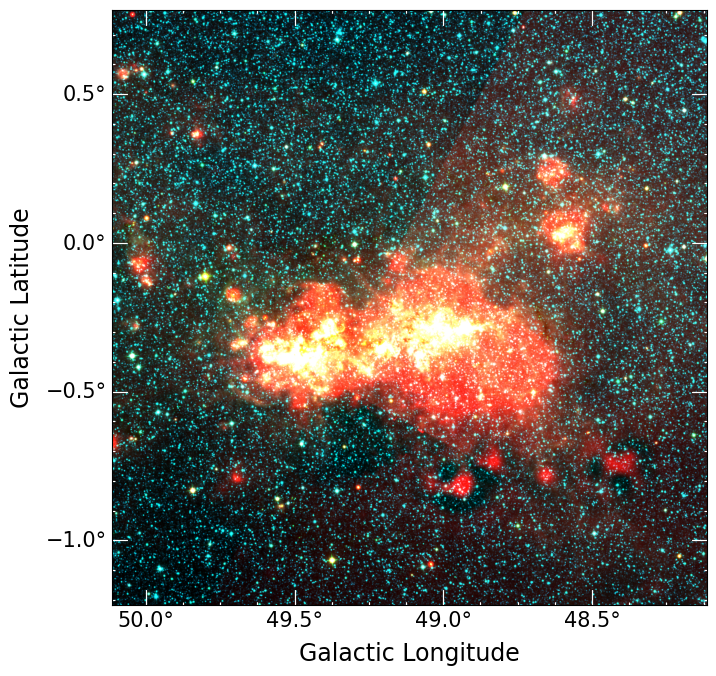}
    \caption{RGB image of the W51 made from WISE W1 (3.4 $\mu$m) in blue, W2 (4.6 $\mu$m) in green, and W4 (22 $\mu$m) in red.
    \label{fig:w51}}
\end{figure}

While these studies have advanced our understanding of variability in nearby low- and intermediate-mass star-forming regions, the variability properties of YSOs in high-mass regions like W51 remain less explored, leaving key questions about their evolution unresolved. This may be due in part to mass sensitivity: W51 primarily hosts intermediate- to high-mass YSOs, and variability mechanisms such as accretion and disk evolution may differ systematically with stellar mass.
To investigate this, we focus on the W51 molecular cloud—a massive, distant star-forming region that provides a valuable laboratory for studying YSO variability in the high-mass regime. 
The W51 molecular cloud, located in the Sagittarius arm about 5.41 kpc from the Sun \citep{sato.2010}, is one of the most massive star-forming regions in the Milky Way. This cloud complex, with a mass exceeding $10^6$ $\Msun$, spans a region about 30 pc across \citep{carpenter.1998} and is divided into three subregions: W51A, W51B, and W51C. These regions host numerous protostars, HII regions, and supernova remnants, making W51 a key site for the study of massive star formation processes (Figure \ref{fig:w51}).

In this study, we performed a systematic analysis of the variability of YSOs in the W51 molecular cloud using NEOWISE mid-infrared data. By studying the variability across different evolutionary stages, we aim to identify the physical mechanisms driving this variability and contribute to a deeper understanding of massive star formation.

\section{Sample and Data}
In this study, the YSO samples in W51 were derived from the catalogs compiled by \citet{Kang.2009} and \citet{Saral.2017}.
\citet{Kang.2009} identified 737 candidate YSOs in W51 and its surroundings using Spitzer GLIMPSE infrared data \citep{Benjamin.2003}, based on infrared color criteria and SED fitting to the YSO model grid of \citet{Robitaille.2006, Robitaille.2007}, from which stellar masses in the range of 0.8–18 $\Msun$ were derived.
\citet{Saral.2017} extended this work by characterizing YSOs in W51 and W43 based on Spitzer color-magnitude diagrams and SED classifications, reporting 302 Class I, 1,178 Class II/transition disk, and approximately 150,000 Class III/Photosphere candidates in W51.

We combined the \citet{Kang.2009} and \citet{Saral.2017} catalogs by cross-matching sources within a 1$''$ radius, resulting in a merged catalog of 160,617 sources. These were categorized into 56 Class 0, 492 Class I, 2,792 Class II, 112 transition disks, and 157,165 Class III/Photosphere objects.

For analysis, we grouped the sources into three main evolutionary categories: Protostar (Class 0/I and flat-spectrum sources), Disk (Class II and transition disks), and PMS+E—a broad category that includes pre-main sequence stars without infrared excess (i.e., Class III) and more evolved sources dominated by stellar photospheres.
Notably, the vast majority of sources ($\sim$157,000) fall into the PMS+E group, reflecting the large population of evolved or non-excess objects in the combined catalog. While this group is dominated by Class III sources from \citet{Saral.2017}, it also includes some sources from \citet{Kang.2009} that lack significant infrared excess.

\label{NEOWISE data}
The Wide-field Infrared Survey Explorer (WISE; \citealt{Wright.2010}), launched in 2009, conducted a full-sky survey and produced the ALLWISE catalog containing photometric data in four bands: 3.4, 4.6, 12, and 22 $\mu$m (W1–W4). Its successor, the NEOWISE Reactivation mission (\citealt{Mainzer.2014}), resumed observations in 2014 and provided time-series photometry in the W1 and W2 bands with a typical cadence of approximately six months until its conclusion in July 2024. We used all 21 observational epochs (about 10.5 yr) available through the end of the mission, enabling long-term monitoring of mid-infrared variability in large samples of YSOs. We note, however, that variations on timescales longer than this time span cannot be detected.

\begin{figure}
    \centering
    \includegraphics[width=\columnwidth]{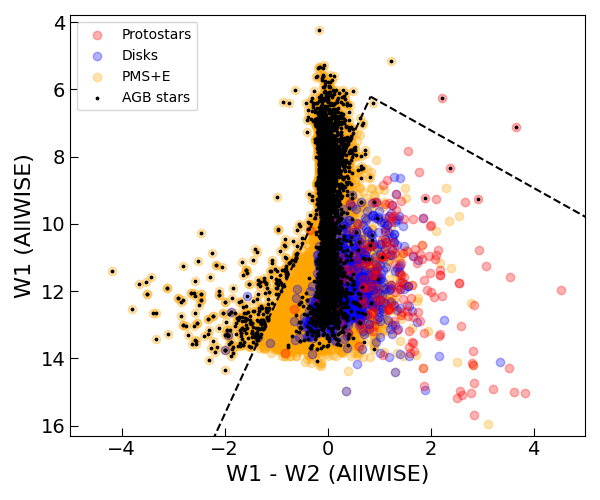}
    \caption{AllWISE color-magnitude diagram. Colored circle symbols indicate the evolutionary stages. Black dots represent sources rejected as contaminants based on the criteria of \citet{Koenig.2014}, including AGB stars, AGNs, and galaxies. The dashed lines indicate the threshold above which sources are excluded as probable AGB stars.
    \label{fig:agb}}
\end{figure}

\label{AllWISE}
Infrared excess from various evolved and extragalactic sources—such as Asymptotic Giant Branch stars (AGBs), AGNs, star-forming galaxies, and classical Be stars—can overlap with YSO signatures in the color-color space \citep{LeeJE.2021}. 
To reduce contamination from these non-YSO sources, we applied the criteria proposed by \citet{Koenig.2014}, which are based on WISE colors and magnitudes. In particular, bright W1 sources with relatively small W1-W2 colors are identified as probable AGB stars and excluded, while additional cuts remove extragalactic contaminants such as AGNs and dusty galaxies. These criteria effectively distinguish YSOs from non-YSOs, yielding a more reliable sample. We removed these contaminants and identified 116,832 sources as likely non-YSOs. Most of these removed sources were located in regions of the color–color diagram corresponding to evolved stars and extragalactic contaminants, primarily falling within the PMS+E category. A small number of sources initially classified as Class II were also excluded, likely due to overlap with AGNs or dusty galaxies. This leaves a refined sample of 43,785 YSO candidates for further analysis (Figure \ref{fig:agb}).

We cross-matched the W51 YSO catalog with the NEOWISE catalog using a 3.2$''$ search radius, corresponding to half the angular resolution of NEOWISE. This process yielded 43,297 matched sources. To ensure photometric reliability, we applied several quality filters. First, each source was required to have detections in more than 12 out of 21 epochs in both W1 and W2. Second, the mean photometric uncertainty at each epoch was required to be less than 0.2 mag.

\begin{figure*}
    \centering
    \includegraphics[width=150mm]{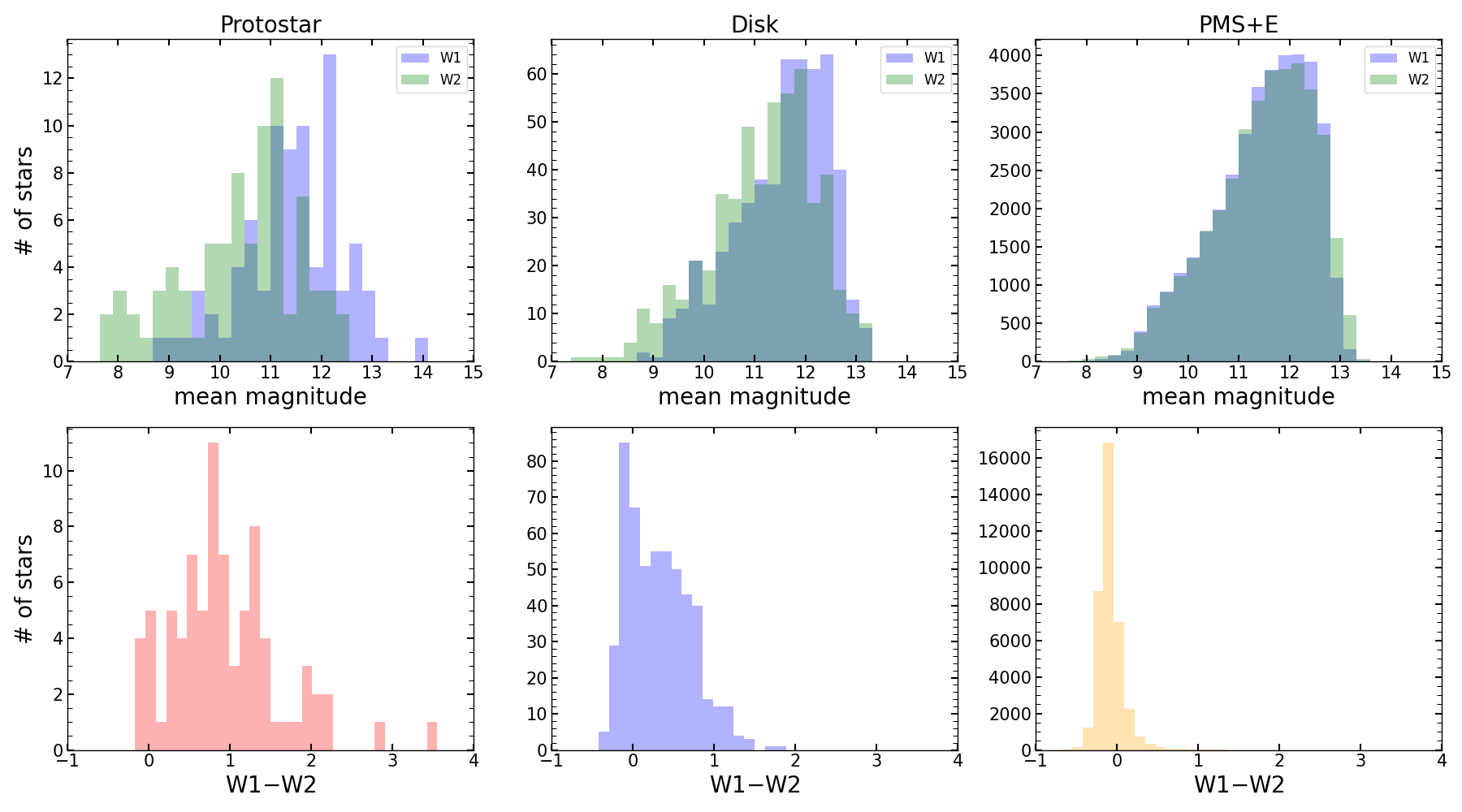}
    \caption{(a) Histograms of the W1 (blue) and W2 (green) mean magnitudes for protostars (left), disk objects (middle), and PMS+E sources (right). (b) Histograms of the W1–W2 color for the same evolutionary classes.
    \label{fig:hist_mag}}
\end{figure*}

After applying the filtering criteria, we obtained a final sample of 38,295 high-quality sources, comprising 81 protostars, 527 disk objects, and 37,687 PMS+E sources.
This dataset enables a statistical investigation of the photometric properties of YSOs across evolutionary stages in the W51 region.
Figure 3 presents the distributions of WISE W1 and W2 mean magnitudes (upper row; a) and W1–W2 color (lower row; b) for each evolutionary class.
The magnitude distributions in (a) show that protostars tend to be fainter than more evolved sources, likely due to greater extinction or intrinsic faintness at mid-infrared wavelengths.
In (b), protostars exhibit a broader and redder color distribution compared to disk and PMS+E sources, consistent with enhanced envelope emission and active accretion in early evolutionary phases. In contrast, PMS+E sources are concentrated at smaller W1–W2 values, reflecting reduced circumstellar material.

\section{Variability in mid-infrared}
Time-domain statistical studies of variability at mid-IR and submm wavelengths \citep{ParkWS.2021,LeeYH.2021,ParkGS.2024,Chen.2025} have established a framework for characterizing YSO variability. In particular, mid-infrared variations observed in the W1 (3.4 $\mu$m) and W2 (4.6 $\mu$m) bands primarily trace emission from hot dust near the sublimation rim and the innermost disk surface, located within a few to several stellar radii of the central star (typically inside 1 au; \citealt{Dullemond.2007,Dullemond.2010}). 
Such regions are highly sensitive to changes in accretion flows, dust structure, and temperature, making mid-IR monitoring a powerful tool for probing the dynamic processes that govern early stellar evolution.

In this study, we analyze the degree and type of variability for our selected targets using the classification method described in \citet{ParkWS.2021}, thus providing a more comprehensive classification of variable types by incorporating both W1 and W2 bands.

\begin{figure*}
    \centering
    \includegraphics[width=\columnwidth]{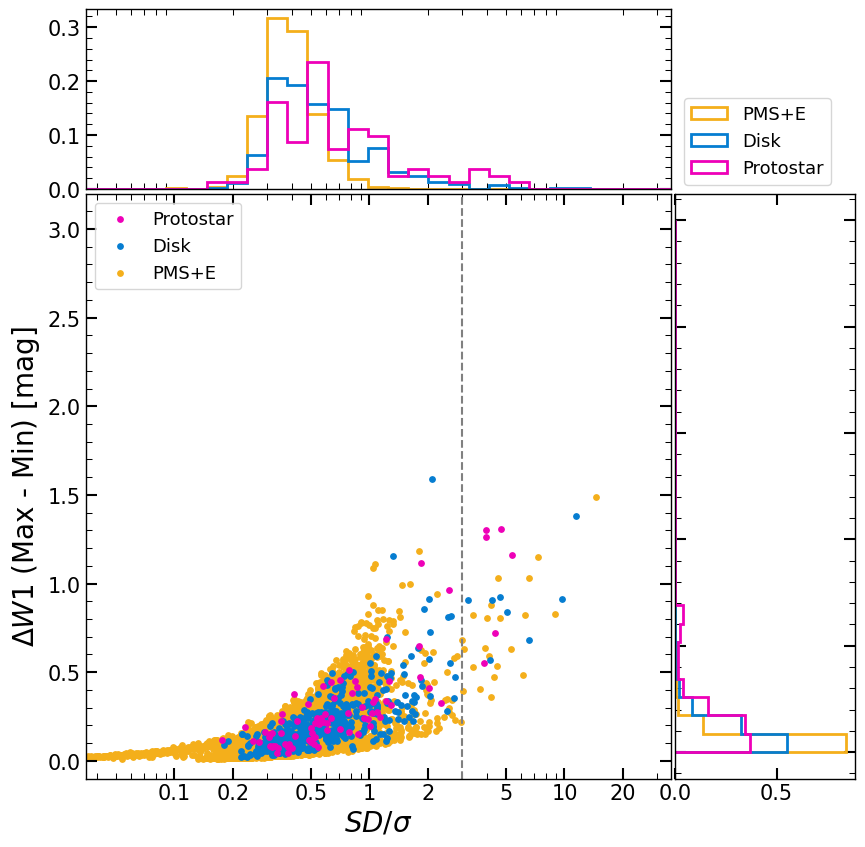}
    \includegraphics[width=\columnwidth]{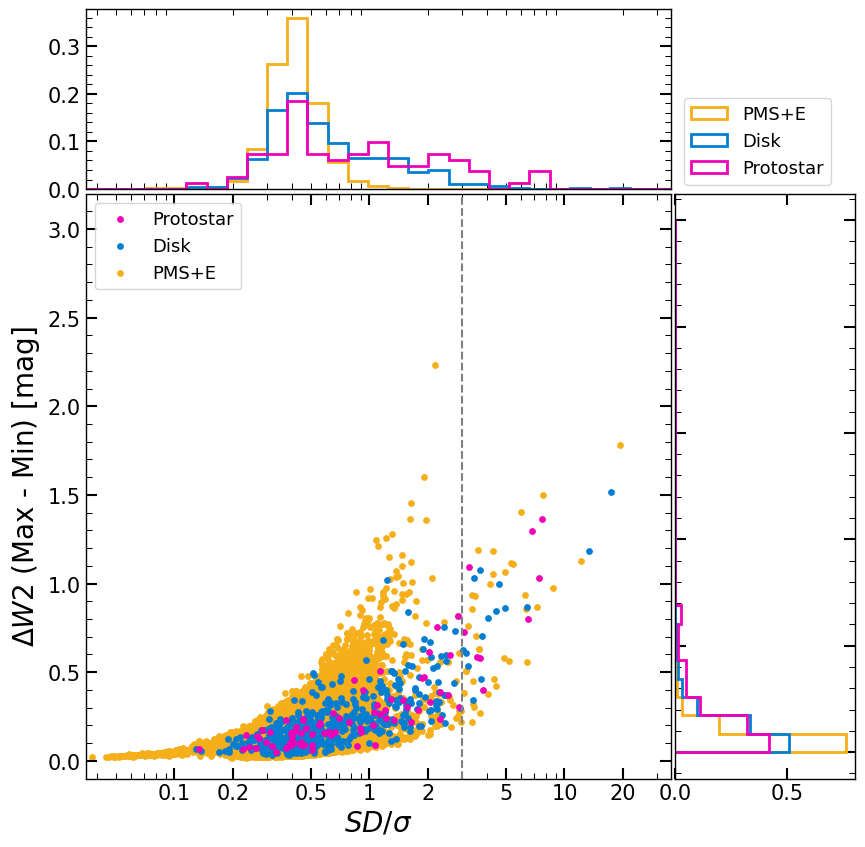}
    \caption{The flux change ($\Delta$W1 and $\Delta$W2) between the maximum and minimum phases as a function of stochasticity (SD/$\sigma$) for all YSOs classified as Protostars, Disk objects, and PMS+E sources. SD represents the standard deviation of fluxes from the light curve, and $\sigma$ denotes the mean flux uncertainty. The vertical dashed line at SD/$\sigma$=3 marks the commonly used threshold for identifying significant variability. Top and right panels show the normalized distributions of SD/$\sigma$ and $\Delta$W, respectively, for each evolutionary stage.
    \label{fig:delW_SD}}
\end{figure*}

\subsection{Variability at Different Stages}
Protostars exhibit broader distributions with higher SD/$\sigma$ and larger $\Delta W$, particularly in the W2 band, indicating stronger variability. The normalized histograms in Figure \ref{fig:delW_SD} show that protostars dominate the higher SD/$\sigma$ regime, while PMS+E sources are more concentrated at lower values. Similarly, protostars have larger $\Delta W$ values compared to disk objects and PMS+E sources, consistent with their early evolutionary stage.

\begin{figure*}
    \centering
    \includegraphics[width=150mm]{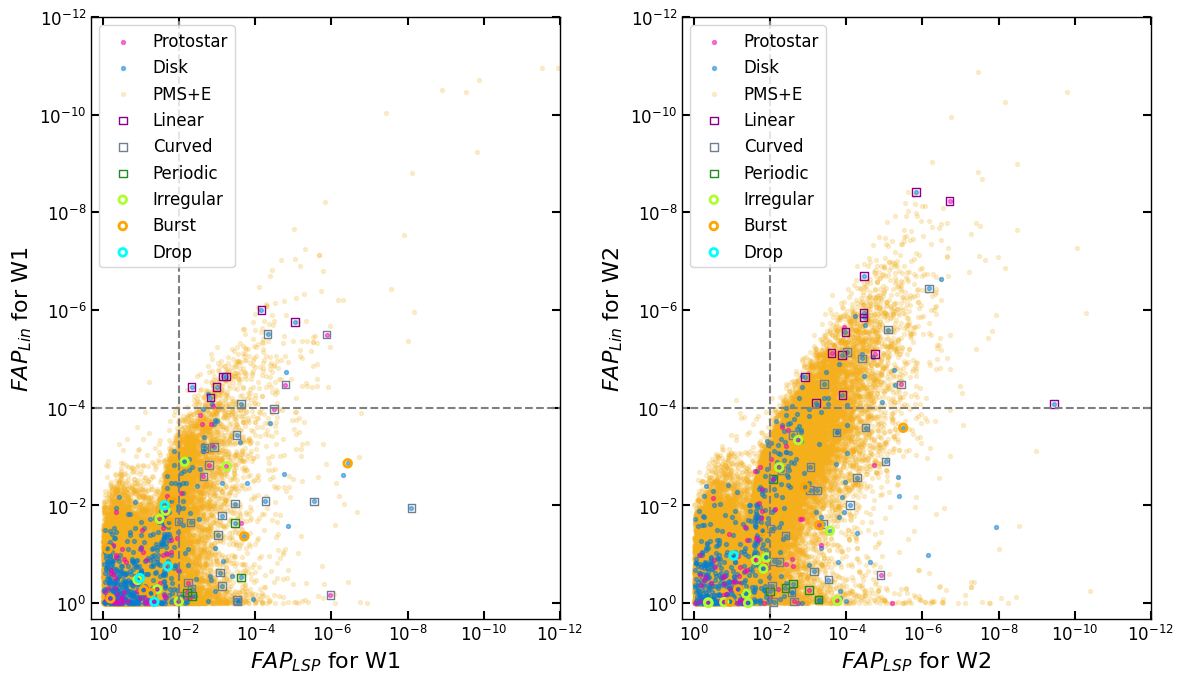}
    \caption{The distribution of linear ($\rm FAP_{Lin}$) and periodic ($\rm FAP_{LSP}$) FAPs for our sample in the W1 (left) and W2 (right). The vertical dashed lines at $\rm FAP_{Lin}$ = $10^{-4}$ separate sources with linear variability, while the horizontal dashed lines at $\rm FAP_{LSP}$ = $10^{-2}$ indicate sources with periodic variability. Small filled symbols represent all sources colored by evolutionary class, while large open markers indicate variability types (linear, curved, periodic, burst, drop, irregular), shown only for classified variable sources among protostars and disks. PMS+E sources, which vastly outnumber the other classes, are plotted in yellow regardless of variability status. 
    \label{fig:FAPs}}
\end{figure*}

We classify variables following \citet{ParkWS.2021} and \citet{ParkGS.2024}. Sources are first required to show significant variability with $SD/\sigma \geq 3$, where $SD$ is the standard deviation of the light curve and $\sigma$ is the mean photometric uncertainty \citep{Johnstone.2018}. The false alarm probability (FAP) quantifies the likelihood of an observed trend or periodic signal emerging purely from random fluctuations rather than reflecting an intrinsic variability. We adopt the analytic formalism of \citet{Baluev.2008}, with further modifications following \citet{LeeYH.2021}, to compute both ${\rm FAP}_{\rm Lin}$ and ${\rm FAP}_{\rm LSP}$.
Secular variables show long-term trends and are subdivided as: (i) \emph{linear}, when a linear fit yields ${\rm FAP}_{\rm Lin} \leq 10^{-4}$; (ii) \emph{curved}, when ${\rm FAP}_{\rm Lin} > 10^{-4}$ but the light curve is better described by a quadratic trend without a significant period; and (iii) \emph{periodic}, when a Lomb–Scargle periodogram \citep{Lomb.1976, Scargle.1989} detects a period $\leq 1900$ days with ${\rm FAP}_{\rm LSP} \leq 10^{-2}$. 
Stochastic variables are identified as (iv) \emph{burst} events, when $(\mathrm{median} - \mathrm{min}) \geq 0.8 \times \Delta W$; or (v) \emph{drop} events, when $(\mathrm{max} - \mathrm{median}) \geq 0.8 \times \Delta W$. Finally, (vi) sources that remain significantly variable but satisfy none of these criteria are classified as \emph{irregular}. 
Figure \ref{fig:FAPs} shows the distributions of ${\rm FAP}_{\rm Lin}$ and ${\rm FAP}_{\rm LSP}$, highlighting the different secular classes.

Table \ref{tab:number} summarizes the variability statistics across evolutionary stages, showing that variability is more common among protostars than disk or PMS+E sources. Younger YSOs tend to show higher SD/$\sigma$ ratios even at similar $\Delta W$ values, reflecting enhanced variability driven by accretion bursts, magnetic activity, or variable extinction. In W51, however, the observed fraction of stochastic variables is notably low, likely due to its greater distance ($\sim$5.4 kpc), reduced photometric precision, and severe source crowding, which limit sensitivity to low-amplitude or short-duration events.

\begin{table}
\caption{Statistics on the number of variables by evolutionary stage and variability type}
\label{tab:number}
\begin{tabular}{lllll} 
\hline
 & Variable type & Protostar & Disk & PMS+E  \\
\hline
W1 & Secular       & 8 (9.9)   & 29  (5.5)  & 167 (0.4) \\  
   & Stochastic    & 6 (7.4)   & 13  (2.5)  & 154 (0.4) \\  
   & Non-variable  & 67        & 485        & 37366 \\
\hline 
W2 & Secular       & 9 (11.1)  & 40 (7.6)   & 219 (0.6) \\
   & Stochastic    & 7 (8.6)   & 12 (2.3)   & 184  (0.5) \\
   & Non-variable  & 65        & 475        & 37284 \\
\hline
total & & 81 & 527 & 37687 \\
\hline
\multicolumn{5}{p{\columnwidth}}{(\%): the fractions relative to the total samples in each evolutionary stage} \\
\end{tabular}
\end{table}

\subsection{Possible Contamination}
The classification of YSOs based on mid-IR variability must take into account potential contamination from evolved stars, such as AGB stars, which often exhibit regular periodic light curves with large amplitudes and periods ranging from hundreds to thousands of days due to stellar pulsations. As shown in Figure \ref{fig:agb}, AGB stars can overlap with YSOs in the color-magnitude diagram, indicating the necessity of careful contamination removal to ensure reliable classification.
Future studies will need to refine contamination mitigation strategies to improve classification accuracy. In particular, distinguishing truly young sources from evolved contaminants such as AGB stars may benefit from incorporating additional observational constraints, including targeted maser surveys \citep{LeeJE.2021} or spectroscopic characterization across multiple wavelengths \citep{Nandakumar.2018}.

As detailed in Table \ref{tab:number}, more than 99\% of PMS+E sources are classified as non-variable, consistently across both W1 and W2 bands. While these sources are not contaminants, their variability characteristics suggest that they include a mix of pre-main sequence stars without inner disks and more evolved main-sequence stars dominated by photospheric emission. In addition, PMS+E is the evolutionary group most susceptible to contamination from background sources such as AGB stars, as discussed earlier.
Given these factors, and our focus on understanding variability in the youngest YSOs, we exclude PMS+E sources from further analysis. This allows us to concentrate on protostars and disk-bearing YSOs, whose mid-infrared variability provides direct insight into physical processes such as accretion, extinction, and disk evolution.

We classified each of the remaining variable sources into secular and stochastic categories according to the criteria described earlier. The resulting numbers of variable sources identified in each category are summarized in Table \ref{tab:variables}, excluding PMS+E sources.
We caution, however, that for both protostars and disks, periodic variability with timescales of ~1 year or longer should be interpreted carefully. Some of these could represent genuine periodic YSOs (e.g., EC 53-like objects), but others may still be unresolved background contaminants such as AGB stars \citep{LeeJE.2021}.

\begin{table}
\caption{Number of variable sources identified in W1 and W2, excluding PMS+E sources.}
\label{tab:variables}
\centering
\begin{tabular}{rlcc}
\hline
Variable & Type & W1 & W2 \\
\hline
& Linear & 7 & 14 \\
Secular & Curved & 26 & 28 \\
& Periodic & 4 & 7 \\
\hline
& Burst & 8 & 6 \\
Stochastic & Drop & 4 & 1 \\
& Irregular & 7 & 12 \\
\hline
\end{tabular}
\end{table}

Detailed variability properties, including variability amplitudes and assigned variability types, for the candidate variable sources classified as Protostars and Disk objects are presented in Table \ref{tab:varlist} (Table \ref{tab:varlist} is available in the online version).

\section{Result and Discussion}
\subsection{Characterization of variability in protostars and disk sources}
\begin{figure*}
\centering
\includegraphics[width=150mm]{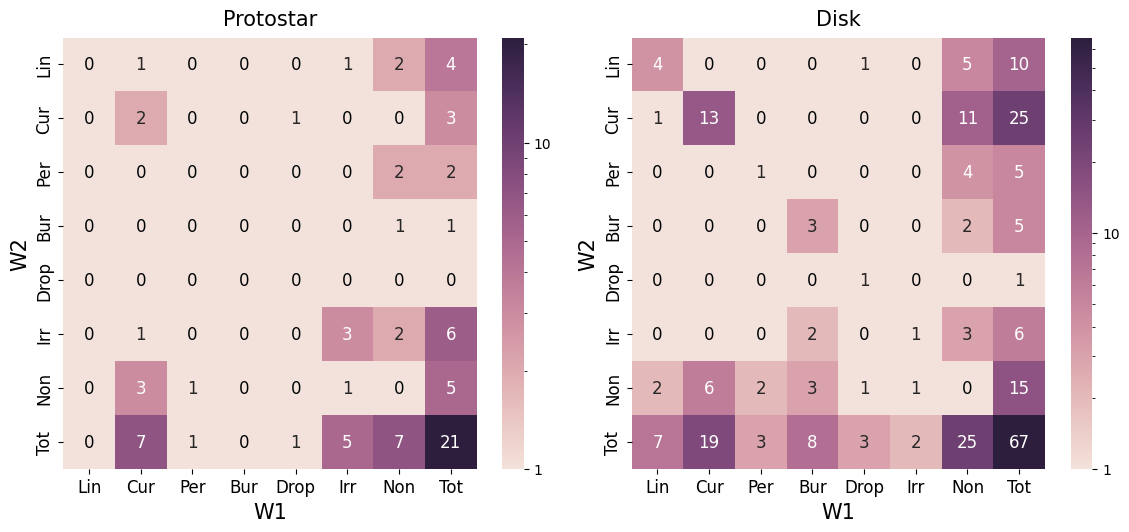}
\caption{Heatmaps illustrating the classification of variability types in the WISE W1 and W2 for Protostars (left) and Disk objects (right). The x-axis represents the variability classification in W1, while the y-axis represents the classification in W2. \label{fig:heatmap}}
\vspace{5mm}
\end{figure*}

Figure \ref{fig:heatmap} presents a two-dimensional count matrix that shows the distribution of variability classifications in the W1 and W2 bands for protostars (left) and disk objects (right).

Protostars predominantly exhibit irregular variability, with 6 out of 21 classified variables (29\%) falling into the irregular category in Figure \ref{fig:heatmap}. Some sources also display periodic (2 objects) and linear (4 objects) types, while burst-type variability is rarely detected (1 object). This distribution suggests a tendency toward stochastic variability in the W2 band, likely influenced by geometric or extinction-related processes. Although the sample size remains limited, this trend is broadly consistent with the mid-infrared variability patterns reported in previous studies \citep{ParkWS.2021}.

Conversely, while the proportion of variable sources decreases from protostars to disks, disk sources exhibit a broader range of variability types. Curved variability is the most common, and burst and periodic types are relatively more frequent compared to protostars. These trends suggest more regulated accretion processes, such as magnetospheric infall and disk-related modulation. After envelope dissipation, accretion occurs through the inner disk, producing more coherent variability signatures. Mid-infrared variability in Class II sources has been interpreted as arising from rotational modulation of asymmetric inner disk structures or from fluctuations in the magnetospheric accretion rate \citep{Fischer.2023}. Consequently, the observed patterns are consistent with a mixture of secular and stochastic processes, depending on the specific disk structure and accretion behavior.

A comparison between W1 and W2 classifications reveals that some sources maintain consistent variability types across both bands, while others exhibit significant discrepancies.
In particular, several sources classified as curved or linear in W1 show irregular or burst-like behavior in W2. Conversely, some burst or periodic variables in W2 appear non-variable or smoothly varying in W1.
This mismatch suggests that the underlying variability mechanisms may be sensitive to the wavelength being probed.
Since W1 is more responsive to hotter inner disk emission and stellar processes, while W2 traces relatively cooler outer disk regions or circumstellar material, the discrepancies likely arise from the spatial and thermal stratification of the emission zones.
Thus, differences in mid-infrared variability classification may reflect variations in the emission mechanisms or wavelength-dependent physical processes, such as dust reprocessing, extinction, or accretion geometry.

\begin{figure*}
\centering
\includegraphics[width=150mm]{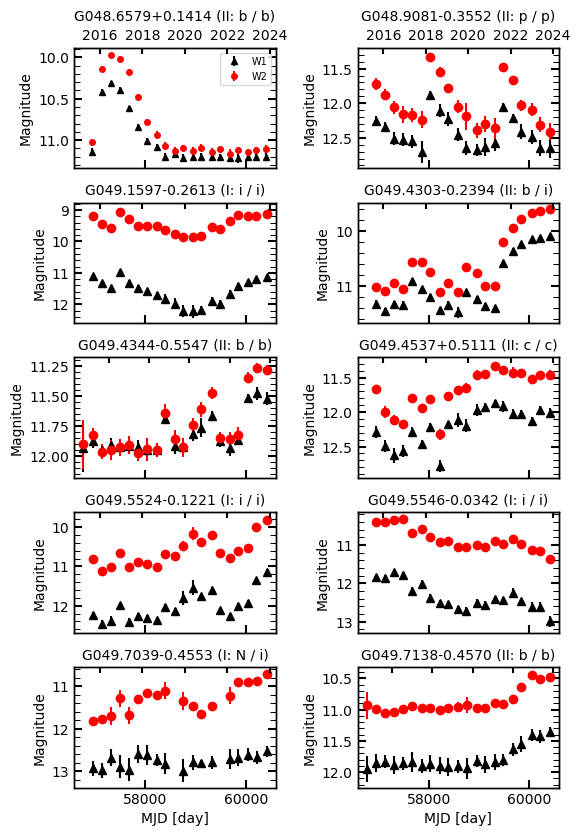}
\caption{Light curves of 10 sources with distinct variability in the W1 (black triangles) and W2 (red circles). The x-axis represents the Modified Julian Date (MJD), while the y-axis shows the magnitude. The top x-axis provides the corresponding year. Error bars indicate photometric uncertainties. The title above each panel lists the source name followed by its evolutionary stage and variability type in each band, in the format (Class: W1 / W2), where W1 and W2 refer to the variability type in the W1 and W2, respectively. The variability types are abbreviated as follows:
c – curved, l – linear, p – periodic, b – burst, i – irregular, N – non-variable.\label{fig:lc}}
\vspace{5mm}
\end{figure*}

To illustrate the diversity of mid-infrared variability, Figure \ref{fig:lc} presents light curves of 10 representative YSOs that exhibit distinct brightness changes in W1, W2, or both bands, with amplitudes approaching 1 mag. These examples were selected to highlight a range of variability types, including curved, burst-like, and irregular behaviors, across different evolutionary stages.

\subsection{Color variability in the CMD}
Color changes in the CMD help us understand what causes brightness changes in young stars. A trend toward bluer colors during brightening can indicate either reduced circumstellar extinction or increased emission from hotter regions, such as the stellar photosphere or the innermost disk edge \citep{Scholz.2013,Parks.2014,Rice.2015,Covey.2021}.

On the other hand, reddening associated with flux increase is not easily explained by extinction effects alone. Instead, it typically reflects an enhanced contribution from warm dust emission, possibly driven by increased accretion or changes in the disk's vertical structure that raise the optical depth along the line-of-sight \citep{Davies.2018,LeeYH.2020,Covey.2021}. Periodic or curved patterns in the CMD may further point to geometrical modulation from orbiting companions or disk asymmetries \citep{Hodapp.2012,Meyer.2019}.

Overall, these mid-infrared changes mostly come from the disk around the star. They reflect how the disk changes shape, moves dust, or heats up, rather than changes on the star itself. This is different from optical variability, which is often caused by spots or magnetic activity on the star’s surface \citep{Bouvier.2013,Cody.2014,Guo.2021}.

\begin{figure*}
\centering
\includegraphics[width=\columnwidth]{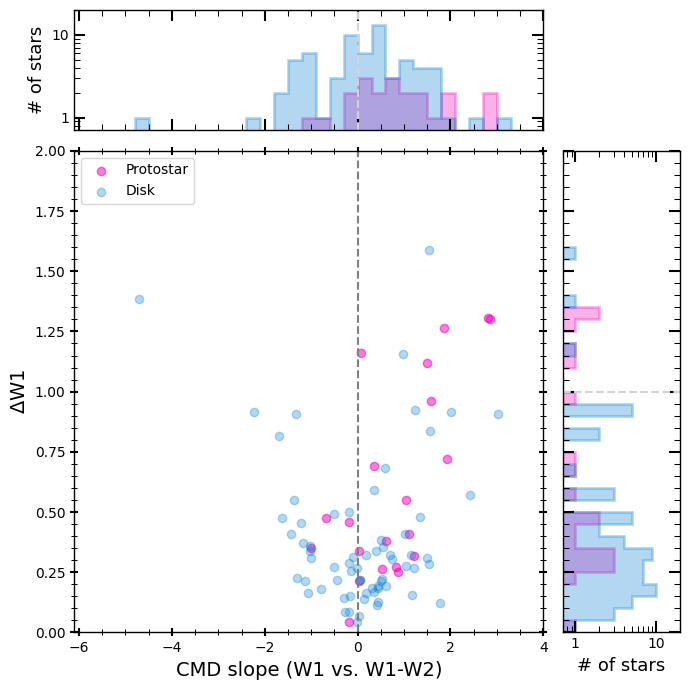}
\includegraphics[width=\columnwidth]{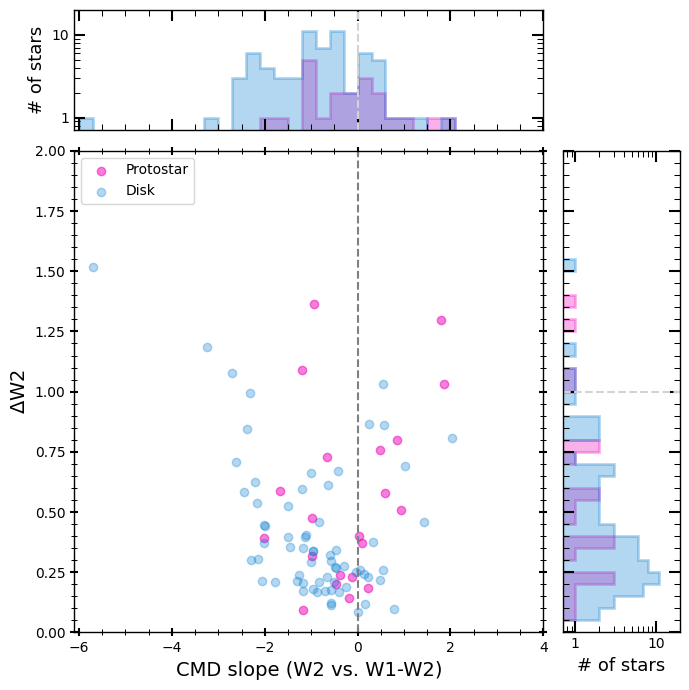}
\caption{Histograms of the CMD slope distributions for Protostar (red) and Disk (blue) objects. The left panel shows the CMD slope for W1 versus (W1-W2), while the right panel displays the CMD slope for W2 versus (W1-W2). The black dotted vertical line represents the zero slope, indicating no color change with brightness variations. The y-axis is on a logarithmic scale to highlight the distribution of stars across a wide range of values. \label{fig:cmd_slope}}
\vspace{5mm}
\end{figure*}

\begin{figure*}
\centering
\includegraphics[width=150mm]{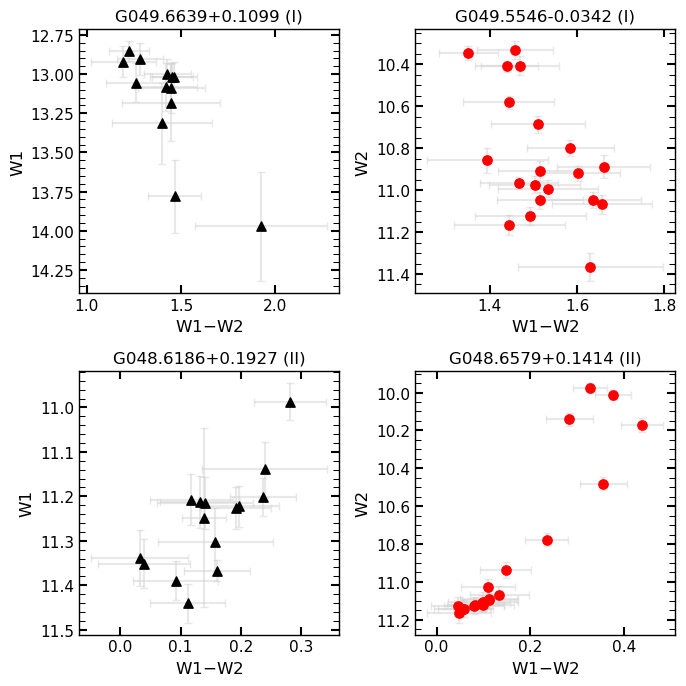}
\caption{Color–magnitude diagrams of four variable YSOs in W51. Black triangles show W1 vs. W1–W2, and red circles show W2 vs. W1–W2. Positive slopes indicate redder colors during brightening, while negative slopes suggest bluer colors. Error bars represent photometric uncertainties. The title of each panel indicates the source name, followed by its evolutionary stage in parentheses. \label{fig:cmd}}
\vspace{5mm}
\end{figure*}

Figure \ref{fig:cmd_slope} presents the distributions of color–magnitude diagram (CMD) slopes as a function of variability amplitude for protostars and disk objects in the W1 (left) and W2 (right). The slope is defined as the change in magnitude (W1 or W2) versus the change in color (W1–W2), where positive slopes indicate redder colors with increasing brightness, and negative slopes imply bluer colors as sources brighten.

To better interpret these trends, we analyzed CMD slopes separately with respect to changes at W1 and W2, as each band probes different physical regions. The W1 traces hotter emission near the stellar surface and inner disk, whereas W2 is more sensitive to cooler disk structures and extinction. This comparison allows us to examine how variability mechanisms differ across wavelengths and evolutionary stages.

In the W1 panel, protostars (81\%) show a clear preference for positive slopes, suggesting that many sources become redder as they brighten, while disk objects (57\%) exhibit a more balanced distribution with a slight tendency toward positive slopes. This behavior may result from enhanced emission from the inner disk, increased hot dust emission, or structural changes in the inner disk wall that contribute additional mid-infrared flux. Protostars, in particular, may also exhibit positive slopes due to the accumulation of disk material increasing the inner disk height and extinction in the early phase of an accretion event \citep{LeeYH.2020, ParkWS.2021}.
To illustrate these trends, Figure \ref{fig:cmd} presents example color–magnitude diagrams for four variable YSOs.

In contrast, the W2 panel shows more complex behavior. Disk objects exhibit a wider range of CMD slopes, with a notable number showing negative values, indicating that these sources become bluer as they brighten. 
This pattern may result from reduced extinction or increased emission from hotter regions near the star, consistent with scenarios involving changes in disk structure or clearing along the line-of-sight. The variety of slopes against W2 variability suggests that mid-infrared variability in this band is shaped by multiple factors, including evolving dust geometry and changes in how stellar radiation is reprocessed. 
On the other hand, positive slopes observed in some sources likely reflect enhanced thermal emission from warm dust, caused by disk heating during periods of stronger accretion.

\subsection{Comparison with variables in other environmental mass star-forming regions}
Our study of the W51 region (at a distance of $\sim$5.4 kpc) provides a unique opportunity to investigate variability in a distant and massive star-forming environment. We identified variable YSOs, allowing for comparison with more nearby and less massive regions such as M17 SWex (at $\sim$1.8 kpc; \citealt{ParkGS.2024}) and the Gould Belt (within $\sim$500 pc; \citealt{ParkWS.2021, LeeSE.2024}).
In the Gould Belt, mid-infrared variability studies have shown that protostars exhibit the highest fraction of variables, reaching up to $\sim$55\%, with large amplitudes and predominantly stochastic light curves driven by episodic accretion and extinction variations \citep{ParkWS.2021}. Long-term secular variability is also common, especially in the earliest evolutionary stages \citep{LeeSE.2024}. In contrast, M17 SWex exhibits lower overall variability fractions (e.g., $\sim$30\% for protostars), though still preserves the trend that younger YSOs show stronger secular variability. The reduced variability in M17 SWex is attributed to its greater distance and photometric sensitivity limits \citep{ParkGS.2024}.

Protostars in W51 also exhibit predominantly stochastic variability, consistent with both the Gould Belt and M17 SWex. However, in W51, we observe a somewhat elevated fraction of stochastic sources accompanied by generally stronger brightness fluctuations compared to nearby regions. While this may reflect more energetic accretion episodes driven by stellar feedback and clustered high-mass star formation, the apparent enhancement could also be influenced by distance effects and increased optical depth in W51’s more embedded environment.

The correlation between W1 and W2 variability types in W51 reveals greater classification discrepancies than in M17 SWex. This may arise from wavelength-dependent factors such as disk temperature gradients, varying optical depths, and reprocessing efficiencies, all of which are likely amplified in W51 due to its higher stellar density and diversity in surrounding disk structures.
Secular variability in W51 is more evenly distributed between W1 and W2 bands, unlike the slight W1 preference noted in M17 SWex. This balance suggests that dust emission and extinction affect both wavelengths comparably in W51, potentially due to more symmetric or optically thick disk structures.
A collective analysis of the Gould Belt and M17 SWex has demonstrated that mid-infrared variability in YSOs is closely linked to evolutionary stage and observational depth. As YSOs evolve from protostars to disks, variability amplitudes typically decline, and variability types become more regulated. W51 exhibits analogous overall trends; however, it demonstrates heightened stochastic behavior, which is presumably indicative of a combination of physical conditions and observational limitations associated with its greater distance, crowding, and energetic environment.

\section{Summary}
In this study, we investigated the mid-infrared variability of YSOs in the W51 star-forming region using NEOWISE photometric data spanning 10 years. 
From a merged catalog of 38,295 mid-IR sources, we identified 81 protostars, 527 disk objects, and 37,687 pre-main sequence and evolved (PMS+E) sources. 
Among these, in the W2 band, 11.1\% of protostars, 7.6\% of disk objects, and 0.6\% of PMS+E sources showed secular variability, while 8.6\% of protostars, 2.3\% of disk objects, and 0.5\% of PMS+E sources exhibited stochastic variability. 
The variability amplitude ($\Delta$W) and stochasticity (SD/$\sigma$) were found to be highest for Protostars, particularly in the W2, suggesting that their variability is influenced by dynamic accretion processes, magnetic activity, and varying extinction. 
Disk objects, although showing lower overall variability fractions than protostars, displayed a wider range of secular variability types—including linear, curved, and periodic patterns—suggesting relatively stable accretion and evolving disk structures. 
The color-magnitude diagram (CMD) analysis revealed distinct trends in color variability across evolutionary stages. 
Protostars tend to show positive CMD slopes, indicating that they become redder as they brighten. This behavior is consistent with increased dust emission or extinction due to the buildup of inner-disk material prior to accretion bursts. Disk objects (57\%) exhibit a more balanced distribution with a slight positive tendency in W1, but in W2 they show a wider range of slopes, with a notable number becoming bluer as they brighten.
This trend likely reflects reduced extinction or hotspot modulation associated with more regulated accretion activity. These contrasting color behaviors underscore the changing dominance of physical mechanisms as YSOs evolve from envelope-dominated protostars to disk-dominated Class II sources.
When compared with other star-forming regions like M17 SWex and the Gould Belt, W51 showed similar trends in variability, such as the prevalence of stochastic variability in protostars and secular variability in disk objects, reflecting consistent physical mechanisms like accretion variability and disk evolution across different environments. 
However, the amplitude and frequency of variability in W51 were generally higher, possibly due to its strong stellar feedback and dense, high-mass environment.

\acknowledgments
This work was supported by the National Research Foundation of Korea (NRF) grant funded by the Korea government (MSIT; grant Nos. 2021R1A2C1011718 and RS-2024-00416859). D.J. is supported by NRC Canada and by an NSERC Discovery Grant. C.C.P. was supported by the National Research Foundation of Korea (NRF) grant funded by the Korean government (MEST) (No. 2019R1A6A1A10073437).

\onecolumn
\begin{table*}[t]
\centering 
\caption{Properties of Variable candidates \label{tab:varlist}}
\begin{tabular}{lcc ccc ll ll} 
\toprule 
ID & Class & Band & N & $\Delta$W & $\rm SD/SD_{fid}$ & $\rm FAP_{LSP}$ & $\rm FAP_{Lin}$ & Sec. & Stoch. \\  
\midrule 
G049.4681-0.0089 & II & W1 & 20 & 0.2 & 1.0 & 2.4E-04 & 8.3E-05 & curved & None  \\ 
                 &    & W2 & 20 & 0.2 & 1.2 & 3.9E-05 & 9.8E-06 & curved & None  \\ 
G049.4035-0.4317 & II & W1 & 17 & 1.2 & 1.3 & 1.1E-01 & 3.1E-01 & None & drop  \\ 
                 &    & W2 & 17 & 0.2 & 0.5 & 2.8E-01 & 6.9E-01 & None & None  \\ 
G048.8184-0.0766 & II & W1 & 20 & 0.2 & 0.7 & 6.0E-04 & 2.3E-05 & linear & None  \\ 
                 &    & W2 & 20 & 0.2 & 1.1 & 3.5E-05 & 1.4E-06 & linear & None  \\ 
G049.2633-0.1759 & II & W1 & 15 & 0.3 & 0.9 & 6.5E-01 & 8.0E-01 & None & burst  \\ 
                 &    & W2 & 15 & 0.2 & 0.5 & 5.8E-02 & 9.1E-02 & None & None  \\ 
G048.6186+0.1927 & II & W1 & 15 & 0.5 & 1.8 & 1.1E-02 & 1.0E-02 & None & None  \\ 
                 &    & W2 & 15 & 0.6 & 3.0 & 4.9E-03 & 1.9E-03 & curved & irregular  \\ 
G048.5943+0.1655 & II & W1 & 20 & 0.2 & 0.7 & 8.8E-02 & 5.4E-01 & None & burst  \\ 
                 &    & W2 & 20 & 0.1 & 0.5 & 9.0E-01 & 1.5E-01 & None & None  \\ 
G048.6579+0.1414 & II & W1 & 20 & 0.9 & 9.7 & 1.9E-02 & 2.2E-02 & None & burst  \\ 
                 &    & W2 & 20 & 1.2 & 13.3 & 1.6E-02 & 3.3E-02 & None & burst  \\ 
G048.8177-0.0766 & II & W1 & 20 & 0.2 & 0.7 & 1.5E-03 & 6.2E-05 & linear & None  \\ 
                 &    & W2 & 20 & 0.2 & 1.1 & 3.5E-05 & 1.2E-06 & linear & None  \\ 
G049.5092+0.2532 & II & W1 & 20 & 0.3 & 0.8 & 1.1E-03 & 3.8E-05 & linear & None  \\ 
                 &    & W2 & 20 & 0.5 & 1.0 & 3.5E-05 & 2.0E-07 & linear & None  \\ 
G048.8184-0.3093 & II & W1 & 20 & 0.2 & 0.5 & 1.0E-01 & 7.1E-02 & None & None  \\ 
                 &    & W2 & 20 & 0.2 & 1.2 & 4.0E-03 & 4.1E-02 & curved & None  \\ 
G048.7741-0.3828 & II & W1 & 20 & 0.0 & 0.3 & 5.4E-01 & 5.3E-01 & None & None  \\ 
                 &    & W2 & 20 & 0.7 & 1.5 & 1.0E-02 & 5.8E-01 & periodic & None  \\ 
G049.5072-0.0249 & II & W1 & 20 & 0.1 & 0.4 & 2.0E-02 & 8.2E-03 & None & None  \\ 
                 &    & W2 & 20 & 0.2 & 1.3 & 3.8E-04 & 3.2E-05 & curved & None  \\ 
G049.6241+0.0279 & II & W1 & 20 & 0.2 & 0.7 & 8.0E-01 & 7.0E-01 & None & None  \\ 
                 &    & W2 & 20 & 0.3 & 1.6 & 5.4E-04 & 8.5E-01 & periodic & None  \\ 
G048.9081-0.3552 & II & W1 & 20 & 0.8 & 2.6 & 4.6E-03 & 7.2E-01 & periodic & None  \\ 
                 &    & W2 & 20 & 1.1 & 3.7 & 2.5E-03 & 4.0E-01 & periodic & irregular  \\ 
G049.1831-0.2111 & II & W1 & 20 & 0.3 & 0.8 & 4.8E-03 & 5.0E-01 & curved & None  \\ 
                 &    & W2 & 20 & 0.4 & 1.2 & 5.9E-04 & 4.9E-03 & curved & None  \\ 
G048.8513-0.4066 & II & W1 & 21 & 0.3 & 0.7 & 1.0E-03 & 1.8E-04 & None & None  \\ 
                 &    & W2 & 21 & 0.2 & 0.9 & 9.6E-05 & 7.3E-06 & curved & None  \\ 
G048.8998-0.4035 & II & W1 & 20 & 0.2 & 0.8 & 5.8E-02 & 1.3E-01 & None & None  \\ 
                 &    & W2 & 20 & 0.3 & 1.4 & 9.0E-03 & 4.5E-01 & curved & None  \\ 
G048.8864-0.4152 & II & W1 & 20 & 0.4 & 1.2 & 7.5E-04 & 4.5E-01 & curved & None  \\ 
                 &    & W2 & 20 & 0.6 & 2.1 & 2.8E-03 & 5.9E-01 & curved & None  \\ 
G049.5688-0.2725 & II & W1 & 20 & 0.4 & 0.9 & 3.4E-04 & 2.3E-02 & periodic & None  \\ 
                 &    & W2 & 20 & 0.3 & 0.6 & 9.9E-04 & 1.5E-04 & None & None  \\ 
G049.4348-0.5204 & II & W1 & 21 & 0.2 & 0.4 & 1.2E-01 & 4.1E-01 & None & None  \\ 
                 &    & W2 & 21 & 0.2 & 0.7 & 1.2E-01 & 9.4E-01 & None & burst  \\ 
G049.1230+0.1173 & II & W1 & 20 & 0.4 & 1.7 & 3.5E-04 & 9.1E-03 & curved & None  \\ 
                 &    & W2 & 20 & 0.5 & 3.2 & 8.0E-05 & 9.9E-03 & curved & irregular  \\ 
G048.8939-0.0851 & II & W1 & 20 & 0.1 & 0.8 & 1.3E-02 & 3.4E-03 & None & None  \\ 
                 &    & W2 & 20 & 0.2 & 0.8 & 9.9E-03 & 3.0E-02 & curved & None  \\ 
G049.2644-0.0816 & II & W1 & 14 & 0.3 & 1.1 & 7.8E-03 & 2.5E-02 & None & None  \\ 
                 &    & W2 & 14 & 0.4 & 1.7 & 8.9E-04 & 1.6E-03 & curved & None  \\ 
G049.3897-0.0152 & II & W1 & 21 & 0.5 & 1.0 & 9.2E-06 & 1.8E-06 & linear & None  \\ 
                 &    & W2 & 21 & 0.4 & 1.8 & 4.1E-03 & 5.0E-04 & None & None  \\ 
G049.6639+0.1099 & I  & W1 & 13 & 1.1 & 1.8 & 4.6E-02 & 9.4E-01 & None & drop  \\ 
                 &    & W2 & 13 & 0.8 & 2.2 & 2.4E-03 & 9.2E-01 & curved & drop  \\ 
\bottomrule 
\end{tabular} 
\tabnote{Only the first 25 rows of the table are shown here. The full table is available in the online supplementary information. This table summarizes the variability characteristics of the selected sources, including secular and stochastic variations in W1 and W2 bands.}  
\end{table*} 
\twocolumn

\bibliography{w51_refs}

\label{lastpage}
\end{document}